\documentclass[%
 reprint,
superscriptaddress,
 amsmath,amssymb,
 aps,
floatfix,
]{revtex4-2}
\usepackage{cancel}
\usepackage[caption=false]{subfig}
\usepackage{graphicx}
\usepackage{dcolumn}
\usepackage{bm}
\usepackage[hidelinks]{hyperref}
\usepackage{xcolor}
\usepackage[normalem]{ulem}
\definecolor{ra}{rgb}{0.8, 0.0, 0.0}

\begin{document}

\preprint{APS/123-QED}

\title{Inverse Renormalization Group in Quantum Field Theory}
\author{Dimitrios Bachtis}
\email{dimitrios.bachtis@swansea.ac.uk}
\affiliation{Department of Mathematics,  Swansea University, Bay Campus, SA1 8EN, Swansea, Wales, UK}
\author{Gert Aarts}
\email{g.aarts@swansea.ac.uk}
\affiliation{Department of Physics, Swansea University, Singleton Campus, SA2 8PP, Swansea, Wales, UK}
\affiliation{European Centre for Theoretical Studies in Nuclear Physics and Related Areas (ECT*) \& Fondazione Bruno Kessler
Strada delle Tabarelle 286, 38123 Villazzano (TN), Italy }

\author{Francesco Di Renzo}
\email{francesco.direnzo@unipr.it}
\affiliation{Dipartimento di Scienze Matematiche, Fisiche e Informatiche, Universit\'a di Parma and INFN, Gruppo Collegato di Parma, I-43100 Parma, Italy}

\author{Biagio Lucini}
\email{b.lucini@swansea.ac.uk}
\affiliation{Department of Mathematics,  Swansea University, Bay Campus, SA1 8EN, Swansea, Wales, UK}%
\affiliation{Swansea Academy of Advanced Computing, Swansea University, Bay Campus, SA1 8EN, Swansea, Wales, UK}

\include{ms.bib}

\date{July 01, 2021}

\begin{abstract}

We propose inverse renormalization group transformations within the context of quantum field theory that produce the appropriate critical fixed point structure, give rise to inverse flows in parameter space, and evade the critical slowing down effect in calculations pertinent to criticality. Given configurations of the two-dimensional $\phi^{4}$ scalar field theory on sizes as small as $V=8^{2}$, we apply the inverse transformations to produce rescaled systems of size up to $V'=512^{2}$ which we utilize to extract two critical exponents. We conclude by discussing how the approach is generally applicable to any method that successfully produces configurations from a statistical ensemble and how it can give novel insights into the structure of the renormalization group.
\end{abstract}

\maketitle

\paragraph*{\label{sec:level1}Introduction.}

Invertibility is a concept that emerges naturally in the mathematical and physical sciences. A simple example of an inverse problem can be defined as follows: given a set of configurations which are sampled in a Monte Carlo simulation, specify the most accurate coupling constants in the underlying Hamiltonian or action of the system that are able to reproduce them.  The problem can be formally expressed as the minimization of a distance metric between two probability distributions under the condition that the model distribution has a dependence on a set of variational parameters, which in the considered case is the set of the coupling constants. The same concept underpins numerous approaches within machine learning. An example arises in quantum field-theoretic machine learning algorithms where arbitrary continuous data can be reproduced based on representations constructed by specifying the optimal values of the coupling constants within algorithms derived from lattice field theories \citep{PhysRevD.103.074510}.

The renormalization group \citep{PhysRevB.4.3174,WILSON197475,RevModPhys.47.773}, which is omnipresent in quantum field theory and statistical physics, is considered to be a non-invertible concept. Scale transformations which construct reduced self-similar representations  of systems necessarily incur some loss of information about the original representation. One should then classify the renormalization group as a semigroup. Nevertheless, the concept of inverse renormalization has been discussed within the context of statistical physics based on systems with simple degrees of freedom, such as the binary Ising model \citep{PhysRevLett.42.859,PhysRevLett.89.275701,PhysRevB.99.075113, PhysRevLett.121.260601}. One then expects that the transition to quantum field theory will give rise to a variety of intricacies, with a central one being the conception of appropriate inverse transformations for systems with continuous degrees of freedom.

\begin{figure}[t]
\includegraphics[width=7.cm]{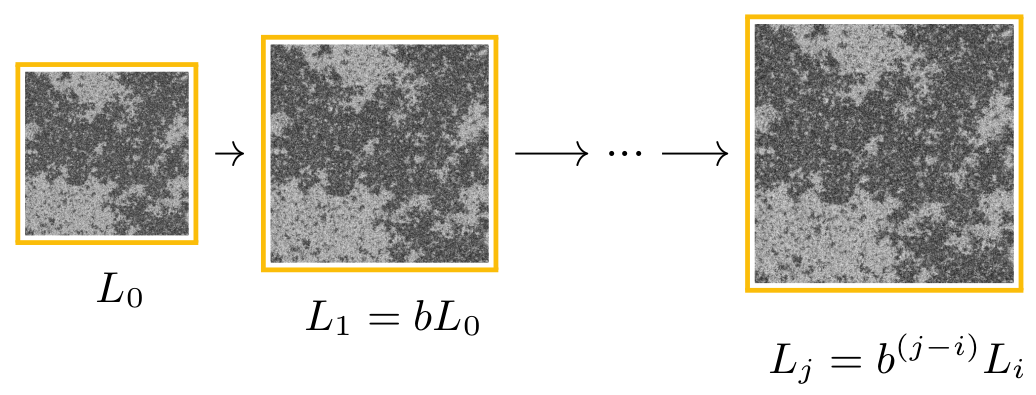}
\caption{\label{fig:1} Illustration of the inverse renormalization group.  Inverse transformations are applied to iteratively increase the size $L$ in each dimension by a factor of $b$, therefore evading the critical slowing down effect.}
\end{figure}

To our knowledge, no inverse renormalization group approach has ever been explored within quantum field theory, despite the fact that computational applications pertinent to the standard renormalization group are numerous, for instance, see Refs.~\citep{PhysRevLett.71.3063,PhysRevD.80.034505,PhysRevLett.108.061601}. The benefits of inverse renormalization would be tremendous: one could apply the transformations iteratively in the vicinity of a phase transition to increase the size of the system and eliminate the critical slowing down effect. Inverse flows in parameter space would then enable the accurate location of the critical fixed point. Furthermore, relations between observables of  the original and the rescaled system could be derived to calculate multiple critical exponents and to obtain complete knowledge of the considered phase transition. Compared to the standard renormalization group, which reduces the size of the system by eliminating degrees of freedom and can hence be applied for only a finite number of steps, inverse renormalization increases the size of the system and can therefore, in principle, be applied for an arbitrary number of steps. 

In this Letter, we propose inverse renormalization group transformations as a means to study phase transitions in quantum field theory.  We explore if the application of inverse transformations is able to iteratively increase the size of the system and if it accurately produces the anticipated flows in parameter space. We then derive expressions between observables of the original and the rescaled system that enable the accurate calculation of multiple critical exponents. The results are illustrated using the second-order phase transition of the two-dimensional $\phi^{4}$ scalar field theory. We conclude by discussing how the approach is generally applicable to any method that produces configurations from a statistical ensemble and how it can give novel insights into the structure of the renormalization group.

\paragraph*{Fundamentals of the inverse renormalization group.}
To construct inverse transformations for systems on graphs or lattices we will devise a set of operations that mimics the inversion of a previously induced transformation. Explicitly, we consider a system of lattice size $L$ in each dimension and apply a renormalization group transformation to reduce its size as
\begin{equation}
L \rightarrow L'=L/b,
\end{equation}
 where $b>1$ is the rescaling factor. All relevant quantities are expressed in terms of lattice units. Our aim is now to learn a set of operations that can mimic the inversion of this transformation:
 \begin{equation}
L' \rightarrow L=bL'.
\end{equation}

The benefit of the approach is that the operations can be iteratively applied to arbitrarily increase the size of the system (see Fig.~\ref{fig:1}). Specifically, if we consider an initial system of size $L$ in each dimension and correlation length $\xi$, then the consecutive applications will produce systems of sizes $L_{0} \rightarrow L_{1}=b L_{0} \rightarrow L_{2}=b L_{1} \rightarrow \ldots$ where the relation describing the increase in the system size at step $j$ is:
\begin{equation}\label{eq:recursivelat}
L_{{j}}=b^{(j-i)} L_{{i}}.
\end{equation}

Here $j>i \geq 0$, and $L_{{0}}=L$. The increase in the lattice size will additionally induce an increase in the correlation length:
\begin{equation}\label{eq:recursivexi}
\xi_{j}=b^{(j-i)} \xi_{i},
\end{equation}
with $\xi_{0}=\xi$. To proceed we introduce the concept of a reduced coupling constant, which is a measure of the distance of a coupling constant $K$ from the critical point $K_{c}$, and which can be defined as:
\begin{equation}
t= \frac{K_{c}-K}{K_{c}}.
\end{equation}

The correlation length $\xi$ arises dynamically in the vicinity of a phase transition and it inherently depends on the distance $t$ from the critical point; in the thermodynamic limit it diverges at $K=K_{c}$. Through the application of iterative transformations which increase the correlation length, each of the rescaled systems will have a different distance $t'$ from the critical point and as a result a different coupling constant $K'$. This is the essence of the renormalization group flow induced in parameter space.

We now consider an intensive observable $O$ in the original system which is a function of the coupling constant $K$. Due to the divergence of the correlation length at the critical point $K_{c}$, the intensive observable quantities $O$ and $O'$ of the original and the rescaled systems will be equal:
\begin{equation}\label{eq:intersect}
O(K_{c})=O'(K_{c}).
\end{equation}

This equation provides a self-consistent manner in locating the critical fixed point: specifically it is the point in parameter space where the observables of the two systems intersect. To locate the critical point it is advisable to compare an original and a rescaled system of the same lattice size to reduce finite size effects \citep{newmanb99}.

Under the condition that observables $O'$ in the rescaled system appear according to the probability distribution of the original system \citep{bachtis2020adding,newmanb99}, we can extrapolate $O'$  along the trajectory of a coupling constant $K$ using histogram reweighting \citep{PhysRevLett.61.2635,bachtis2020adding,bachtis2020extending,PhysRevE.102.053306}, while relying on the action $S$ of the original system:
\begin{equation}
\langle O' \rangle= \frac{\sum_{l=1}^{N} O_{\sigma_{l}}' \exp[-(K_{m}-K_{m}^{(0)}) S_{\sigma_{l}}^{(m)})]}{\sum_{l=1}^{N} \exp[-(K_{m}-K_{m}^{(0)}) S_{\sigma_{l}}^{(m)})]},
\end{equation}
where $\sigma_{l}$ is a configuration of the system, $N$ is the number of samples, and the action $S=\sum_{m} K_{m}^{(0)} S^{(m)}$ of the original system is expressed as a sum over products of coupling constants $K_{m}^{(0)}$ and their corresponding action terms $S^{(m)}$. In the example considered above, the discussed histogram reweighting approach considers strictly the extrapolation of exclusively one coupling constant $K$ in parameter space.

A critical exponent that characterizes a phase transition is the exponent $\beta$ which couples to the magnetization $m_{i} \sim |t_{i}|^{\beta}$ and $m_{j} \sim |t_{j}|^{\beta}$, and which can be equivalently expressed in terms of the correlation length as $m_{i} \sim \xi_{i}^{-\beta/\nu}$ and $m_{j} \sim \xi_{j}^{-\beta/\nu}$, where $\nu$ is the exponent that governs the divergence of $\xi$. By dividing, substituting,  and taking the natural logarithm of the expressions we arrive at the relation:
\begin{equation}
\frac{\beta}{\nu}= -\frac{\ln \frac{m_{j}}{m_{i}}}{\ln \frac{\xi_{j}}{\xi_{i}}} =- \frac{\ln \frac{m_{j}}{m_{i}}}{(j-i)\ln b}.
\end{equation}

The above expression can be redefined to be suitable for a finite system using l'H$\mathrm{\hat{o}}$pital's rule \citep{newmanb99}, arriving at the expression:
\begin{equation}\label{eq:beta}
\frac{\beta}{\nu}= -\frac{\ln \frac{dm_{j}}{dm_{i}} \big|_{K_{c}}}{\ln \frac{\xi_{j}}{\xi_{i}}}=-\frac{\ln \frac{dm_{j}}{dm_{i}} \big|_{K_{c}}}{(j-i)\ln b}.
\end{equation}

Following a similar procedure for the magnetic susceptibility $\chi_{i} \sim |t_{i}|^{-\gamma}$ and $\chi_{j} \sim |t_{j}|^{-\gamma}$ we obtain:
\begin{equation}\label{eq:gamma}
\frac{\gamma}{\nu}= \frac{\ln \frac{d\chi_{j}}{d\chi_{i}} \big|_{K_{c}}}{\ln \frac{\xi_{j}}{\xi_{i}}}=\frac{\ln \frac{d\chi_{j}}{d\chi_{i}} \big|_{K_{c}}}{(j-i) \ln b}.
\end{equation}

Using the above equations and the renormalization group, one can calculate critical exponents through numerical derivatives of the observables $m$ and $\chi$ in the vicinity of the critical point $K_{c}$.

\paragraph*{Inverse renormalization in the $\phi^{4}$ theory.}
\begin{figure}[t]
\includegraphics[width=8.6cm]{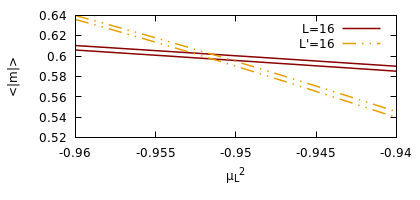}
\caption{\label{fig:3} Absolute value of the magnetization versus the dimensionless squared mass. $L'$ denotes a system produced with the standard renormalization group approach which reduces the size of the system. The region bounded by the lines denotes the statistical uncertainty.}
\end{figure}

We consider the discretized two-dimensional $\phi^{4}$ scalar field theory on a square lattice  with the lattice action \citep{PhysRevD.103.074510}:
\begin{equation}
S=- \kappa_{L} \sum_{\langle ij \rangle} \phi_{i} \phi_{j} + \frac{(\mu_{L}^{2}+4 \kappa_{L})}{2} \sum_{i} \phi_{i}^{2} + \frac{\lambda_{L}}{4} \sum_{i} \phi_{i}^{4}.
\end{equation}
Here $\kappa_{L}$, $\mu_{L}^{2}$, $\lambda_{L}$ are dimensionless parameters. The system undergoes a second-order phase transition between a symmetric and broken-symmetry phase for specific values of $\mu^{2}_{L} < 0$ when  $\lambda_{L}>0$ and $\kappa_{L}>0$ \citep{Milchev1986}. We will consider the case $\lambda_{L}=0.7$, $\kappa_{L}=1$ and vary the coupling constant $\mu^{2}_{L} \equiv K$. We simulate the system using a combination of the Metropolis and Wolff algorithms \citep{PhysRevLett.62.1087,PhysRevD.79.056008,PhysRevD.58.076003, PhysRevLett.62.361}, and the errors are calculated with a binning analysis using $10^{4}$ configurations in $10$ separate bins. Observables of interest are the magnetization $M=|\sum_{i} \phi_{i}|$, and the magnetic susceptibility $\chi= (1/V) (\langle M^{2} \rangle- \langle M \rangle^{2})$. We denote as $m=(1/V)M$ the intensive magnetization which is normalized by the size of the system $V=L \times L$.

Starting from a $\phi^{4}$ theory with lattice size $L=32$ in each dimension, we first apply a standard renormalization group transformation with $b=2$ on configurations sampled at $\mu_{L}^{2} = -0.9515$ in the vicinity of the phase transition  to produce a rescaled system with size $L'=16$.  Specifically the transformation consists of separating the system in blocks of size $b \times b$, where the degrees of freedom are summed within each block. If the sum is positive or negative then we select the rescaled degree of freedom as the mean of the positive or negative degrees of freedom within the block, respectively. Since the lattice size is halved the correlation length will be reduced similarly, $\xi'=\xi/2$. The emergent renormalization group flow then drives the system away from the critical point towards either the broken-symmetry or the symmetric phase, depending on where the system was initially positioned in. This implies that if the original system had a certain magnetization $m$ then the rescaled system will have magnetization $m'>m$ ($m'<m$) if it was initially in the broken-symmetry (symmetric) phase. The results, obtained with the use of histogram reweighting, are depicted in Fig.~\ref{fig:3} where the standard renormalization group flow and a critical fixed point have emerged. 

Every successful standard renormalization group transformation encodes important information. First, that the original and the rescaled systems are an accurate representation of the same physical model. Second, that configurations of the rescaled system follow the probability distribution of the original system, and, third, that a critical fixed point structure exists at criticality. We have verified, through the obtained results, that the standard renormalization group transformation, implemented as above, satisfies these conditions. By learning how to mimic the inversion of this transformation we anticipate that the same conditions will additionally be satisfied on the inverse transformation. The inverse transformation can then be iteratively applied to arbitrarily increase the size of the system.

To mimic the inversion of a transformation we will rely on the application of a set of transposed convolutions~\citep{dumoulin2018guide}, which can be easily implemented, for instance, via the Keras library~\citep{chollet2015keras}.  Details can be found in the Supplemental Material \footnote{See Supplemental Material at [URL will be inserted by publisher], which includes Refs~[22-23], for details about the inverse renormalization group transformation}. The input system to the transposed convolutions is the rescaled system with size $L'=16$ and the output is a system with size $L=32$ which is equal to the original.  We remark that the inverse transformation is not anticipated to be a perfect inversion of the original one. Importantly,  the set of transformations have no dependence on the size of the system and can therefore be applied to any arbitrary size $L$.

\begin{figure}[t]
\includegraphics[width=8.6cm]{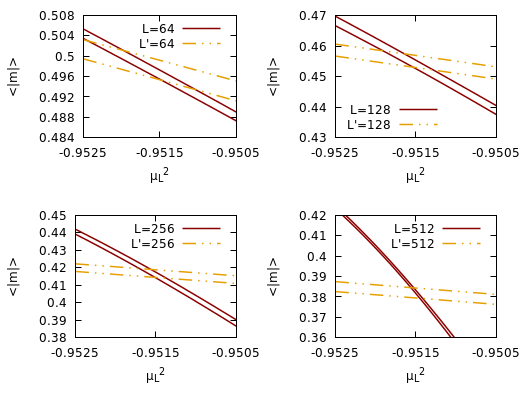}
\caption{\label{fig:4} Absolute value of the magnetization versus the dimensionless squared mass. $L'$ denotes a system produced with the inverse renormalization group approach which increases the size of the system. The region bounded by the lines denotes the statistical uncertainty.}
\end{figure}

We will now apply the inverse transformations to iteratively increase the lattice size by a factor of $b=2$ through the relation of Eq.~\ref{eq:recursivelat}. We anticipate that the iterative increase in the lattice size will also equally increase the correlation length (see Eq.~\ref{eq:recursivexi}), under the condition that there exists some finite correlation length present in the initial configurations, therefore driving the system towards the critical point irrespective of the phase that it is initially positioned in. This implies that if the original system had magnetization $m$ then the rescaled system will have magnetization $m'<m$ ($m'>m$) if it was initially in the broken-symmetry (symmetric) phase, respectively. The results are depicted in Fig.~\ref{fig:4}. We observe, based on the intersection of observables (see Eq.~\ref{eq:intersect}), that the critical fixed point agrees with the expected values of $\mu^{2}_{c}=-0.95151(25)$\cite{PhysRevD.79.056008}, $\mu^{2}_{c}=-0.9516(8)$\citep{PhysRevD.58.076003}, and that the anticipated behaviour of the inverse flows in parameter space has emerged. The previous results, which relied on a comparison of the rescaled versus the original system at the same lattice size, served as a proof-of-principle demonstration to establish the inverse renormalization group approach. In fact, this comparison is neither needed nor desired because it requires the simulation of the original system at all lattice sizes and is therefore hindered by the critical slowing down effect. 

\begin{figure}[b]
\includegraphics[width=8.6cm]{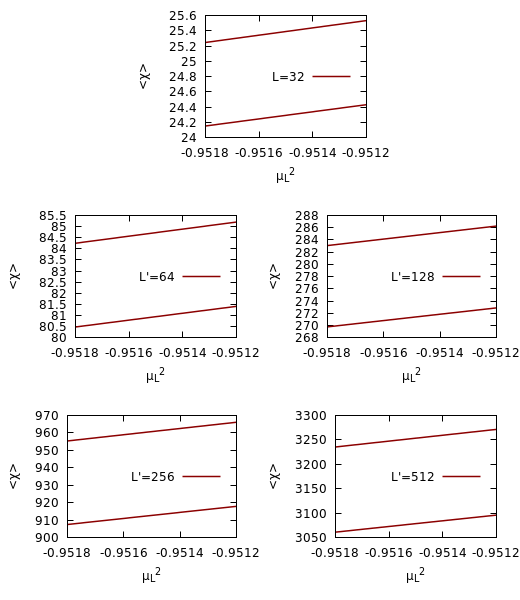}
\caption{\label{fig:5}Magnetic susceptibility $\chi$ versus the dimensionless squared mass. The region bounded by the lines denotes the statistical uncertainty.}
\end{figure}

\begin{table*}[t]
\caption{\label{tab:data} Values of the critical exponents $\gamma/\nu$ and $\beta/\nu$. The original system has lattice size $L=32$ in each dimension and its action has coupling constants $\mu_{L}^{2}=-0.9515$, $\lambda_{L}=0.7$, $\kappa_{L}=1$. The rescaled systems are obtained through inverse renormalization group transformations.}
\begin{ruledtabular}
\begin{tabular}{c|c|c|c|c|c|c|c|c|c|c|}
$L_{i}/L_{j}$ & 32/64 & 32/128 & 32/256 & 32/512 & 64/128 & 64/256 & 64/512 & 128/256 & 128/512 & 256/512 \\
\hline
$\gamma/\nu$  & 1.735(5) & 1.738(5)  & 1.741(5)  & 1.742(5)  & 1.742(5)  & 1.744(5)  & 1.744(5)  & 1.745(5)  & 1.745(5) & 1.746(5) \\
$\beta/\nu$ & 0.132(2) & 0.130(2) & 0.128(2) & 0.128(2) & 0.128(2) & 0.127(2) & 0.127(2) & 0.126(2) & 0.126(2) & 0.126(2) \\
\end{tabular}
\end{ruledtabular}
\end{table*}

\begin{table*}[t]
\caption{\label{tab:data2} Values of the critical exponents $\gamma/\nu$ and $\beta/\nu$. The original system has lattice size $L=8$ in each dimension and its action has coupling constants $\mu_{L}^{2}=-1.2723$, $\lambda_{L}=1$, $\kappa_{L}=1$. The rescaled systems are obtained through inverse renormalization group transformations.}
\begin{ruledtabular}
\begin{tabular}{c|c|c|c|c|c|c|c|c|c|c|c|}
$L_{i}/L_{j}$ & 8/16 & 8/32 & 8/64 & 8/128 & 8/256 & 8/512 & 16/32 & 16/64 & 16/128 & 16/256 & 16/512 \\
\hline
$\gamma/\nu$  & 1.694(6) & 1.708(6)  & 1.717(6)  & 1.723(6)  & 1.727(6)  & 1.730(6)  & 1.721(6)  & 1.728(6)  & 1.732(6) & 1.735(6) & 1.737(6) \\
$\beta/\nu$ & 0.154(2) & 0.147(2) & 0.142(2) & 0.139(2) & 0.137(2) & 0.135(2) & 0.140(2) & 0.136(2) & 0.134(2) & 0.132(2) & 0.131(2) \\
\end{tabular}
\begin{tabular}{c|c|c|c|c|c|c|c|c|c|c|}
$L_{i}/L_{j}$ & 32/64 & 32/128 & 32/256 & 32/512 & 64/128 & 64/256 & 64/512 & 128/256 & 128/512 & 256/512 \\
\hline
$\gamma/\nu$  & 1.735(6) & 1.738(6)  & 1.740(6)  & 1.740(6)  & 1.741(6)  & 1.742(6)  & 1.742(7)  & 1.743(6)  & 1.743(7) & 1.743(7) \\
$\beta/\nu$ & 0.133(2) & 0.131(2) & 0.130(2) & 0.129(2) & 0.129(2) & 0.129(2) & 0.128(2) & 0.128(2) & 0.127(2) & 0.127(2) \\
\end{tabular}

\end{ruledtabular}
\end{table*}

The critical slowing down effect can be entirely avoided in calculations pertinent to criticality through the use of Eqs.~\ref{eq:beta} and~\ref{eq:gamma}. Based on the original system with $L_{0}=32$ we obtain with the inverse transformations a set of rescaled systems $L_{j}=64,128,256,512$, from which we calculate two critical exponents through a numerical derivative of the magnetization and the magnetic susceptibility. Since the method does not require any additional simulation  in the vicinity of the phase transition, other than the one at $L_{0}=32$, no critical slowing down effect emerges.  In addition, it is possible to compare two rescaled systems, for instance the ones with $L_{3}=256$ versus $L_{4}=512$ to further increase the accuracy of the results as the comparison between larger lattices will substantially diminish finite size effects. 

Results for all possible sets of systems are provided in Table~\ref{tab:data} and the magnetic susceptibility for the rescaled systems is depicted in Fig.~\ref{fig:5}. We calculate the critical exponents based on the same range of coupling constants $ -0.9516 \leq \mu_{L}^{2} \leq -0.9514$, to guarantee consistency in the results.  We observe that there is a clear convergence towards the expected values of $\gamma/\nu=7/4=1.75$ and $\beta/\nu=1/8=0.125$ of the Ising universality class as the comparison between systems is conducted on larger lattice sizes, therefore diminishing finite size effects. 

To demonstrate that the learned set of transformations is generally applicable to different lattice sizes, as well as to different points in parameter space, we apply the inverse renormalization group to configurations obtained at a different set of coupling constants along the critical line (see Ref.~\cite{PhysRevD.79.056008}). Specifically, we simulate the $\phi^{4}$ scalar field theory with lattice size $L_{0}=8$ in each dimension and a set of coupling constants $\kappa_{L}=1$, $\mu_{L}^{2}=-1.2723$ and $\lambda_{L}=1$. We then apply the inverse transformations to obtain systems of lattice size up to $L=512$, from which we calculate the critical exponents. The results are depicted in Table~\ref{tab:data2}, where we observe a convergence towards the anticipated values, therefore verifying that the method is applicable to different lattice volumes and to phase transitions that occur in different regions of parameter space. 

In summary, through the use of inverse renormalization group transformations we were able to iteratively increase the size of the system in absence of the critical slowing down effect and to obtain two critical exponents of the second-order phase transition. It is intriguing that the combination of the probabilistic perspective and the inverse renormalization is able to produce extrapolations of observables for the iteratively increasing lattice sizes $L_{j}=16, 32, 64, 128, 256, 512$ given exclusively one Monte Carlo simulation obtained at one point in parameter space for lattice size $L_{0}=8$ or $L_{0}=32$. This information  would have not been otherwise accessible and could have been previously obtained only through the use of computationally demanding simulations conducted directly at the specific lattice sizes $L_{j}$.

\paragraph*{Conclusions.}
We have shown that inverse renormalization group transformations emerge as an approach  within quantum field theory which is able to evade the critical slowing down effect in numerical calculations pertinent to criticality. Specifically, using the two-dimensional $\phi^{4}$ scalar field theory of lattice size $L_{0}=8$ or $L_{0}=32$ in each dimension, we applied the inverse transformations to iteratively increase the size of the system to $L'=512$, without the need to conduct additional simulations, and we observed the induced renormalization group flow in parameter space. The approach enables the accurate extraction of the critical exponents for the magnetization and the magnetic susceptibility using exclusively configurations produced from the inverse transformations.

Numerous research directions can be envisaged. Quantum field-theoretic machine learning algorithms \citep{PhysRevD.103.074510} can be implemented to learn the appropriate coupling constants of the rescaled systems allowing for complete physical interpretability of the results. The structure of the inverse renormalization group transformations and the emergent flows could then be understood fully.  Furthermore, the extraction of additional critical exponents can be achieved by introducing terms which induce symmetry-breaking in the original system. These terms could be extrapolated to the iteratively rescaled ones through the use of histogram reweighting which is agnostic to the form of the underlying original and renormalized action \citep{bachtis2020adding}. Furthermore, possible extensions of the inverse renormalization group to multiscale methods which implement real-space transformations could be explored \citep{PhysRevD.92.114516}. In addition, one could construct the linearized renormalization group transformation matrix \citep{PhysRevLett.42.859}, using the rescaled configurations to extract the relevant operators. Computational investigations of the renormalization group have been applied in a diverse range of quantum field theories \citep{PhysRevLett.108.061601,PhysRevLett.71.3063,PhysRevD.80.034505}, including quantum chromodynamics, and inverse transformations within these systems are therefore open to explore.  Finally, the method only requires one set of configurations in the vicinity of the phase transition and it is therefore generally applicable to any approach that successfully samples configurations from a statistical ensemble. 

In conclusion, the inverse renormalization group, an approach that successfully evades the critical slowing down effect which has hindered numerical simulations of systems that undergo phase transitions since their initial conception,  is a vastly unexplored concept within quantum field theory, and further exploration could potentially yield novel mathematical and physical insights into the structure of the renormalization group, thereby paving the way for a deeper understanding of a concept ubiquitous in physics. 

\textbf{ Note added}: While this work was being submitted, we became aware of the investigations reported in Ref.~\citep{Shiina2021}, which proposes related ideas applied to discrete spin systems.

\paragraph*{\label{sec:level5}Acknowledgements.}
The authors received funding from the European Research Council (ERC) under the European Union's Horizon 2020 research and innovation programme under grant agreement No 813942. The work of GA and BL has been supported in part by the UKRI Science and Technology Facilities Council (STFC) Consolidated Grant ST/T000813/1. The work of BL is further supported in part by the Royal Society Wolfson Research Merit Award WM170010 and by the Leverhulme Foundation Research Fellowship RF-2020-461\textbackslash 9. FDR acknowledges partial support from I.N.F.N. under the research project i.s. QCDLAT. Numerical simulations have been performed on the Swansea SUNBIRD system. This  system is part of the Supercomputing Wales project, which is part-funded by the European Regional Development Fund (ERDF) via Welsh Government.

\bibliography{ms}
\end{document}